\documentclass{csmagclass}														
\usepackage{graphicx}
\usepackage{subfigure}
\usepackage{amsmath}
\usepackage{amssymb}	
\begin{document}		
\headings																															 %

\title{{ Emergence of a Skyrmion Phase in a Frustrated Heisenberg Antiferromagnet with Dzyaloshinskii-Moriya Interaction}}
\author[1]{{M. MOHYLNA}}
\author[1]{{ M. \v{Z}UKOVI\v{C}\thanks{Corresponding author: milan.zukovic@upjs.sk}}}
\affil[1]{{Institute of Physics, Faculty of Science, P. J. Šafárik University, Park Angelinum 9, 041 54 Košice, Slovakia}}

\maketitle

\begin{Abs}
We study formation of a periodical structure of quasiparticle-like magnetic configurations with non-trivial topological charge, known as a magnetic skyrmion phase, on a frustrated triangular lattice antiferromagnetic Heisenberg model with Dzyaloshinskii-Moriya interaction (DMI) by means of Monte Carlo simulations. The existence of such a phase, formed on the three interpenetrating sublattices, has been demonstrated for a sufficiently large strength of DMI. The goal of the present investigation is to establish the minimum values of both the DM interaction strength as well as the magnetic field intensity at which such a skyrmion phase can exist at very low temperatures (close to zero). We find that the skyrmion phase emerges above the DMI parameter value of $D = 0.2$ and persists within the field intensity $2 \lesssim h < 4.5$.

\end{Abs}
\section{Introduction}
In recent years topological solitons have attracted a lot of attention due to their unusual properties and a wide variety of applications. Being introduced in the 60th by T. Skyrme in attempt to explain the stability of baryons in the framework of non-linear $\sigma$ model~\cite{ref1}, they have since been adopted in many fields of physics as a powerful tool for universal description of non-linear phenomena. From the point of view of both fundamental research and practical implementation of particular interest are topologically protected structures emerging in magnetic materials. Some of them, like magnetic domain walls, are known for almost a century~\cite{ref2} and some others, like skyrmions, are being currently actively investigated. One of the main reasons for the growth of their popularity is the topological stability they exhibit: each type of a soliton is ascribed a certain property called topological charge and topological configurations with different charges cannot be continuously transformed into each other due to an infinite energy barrier~\cite{ref3}. This means, in particular, that configurations with non-zero topological number cannot be reduced to a ferromagnetic state, which corresponds to a trivial state in terms of topology. 

A periodical skyrmion lattice structure in magnetic materials was theoretically predicted in 80th~\cite{ref4}, but it wasn't until 2009 when the solid proof of the existence of skyrmion phase in B20 bulk alloys was provided by means of small angle neutron scattering~\cite{ref5}. However, for ferromagnetic bulk materials the region in $T-h$ space, where such phase is stabilized, turns out to be rather narrow and is close to the Curie temperature. Further research has shown, that reduction of a sample's thickness may notably increase the window where skyrmions exist, broadening it even to zero temperatures~\cite{ref6}. Various skyrmion-based devices have been proposed since then, owning to the fact that they are highly stable, mobile and have a size up to a couple of hundreds of nm. Skyrmions are sensitive to and can be created by spin-polarized currents of low densities, magnon currents in insulators, electric fields and temperature gradient, which makes them suitable candidates for implementation in different magnetic memory storages, logic and microwave devices. However, skyrmions in ferromagnets have certain drawbacks and limitations, and thus the possibility  of using antiferromagnetic materials (AFM) have recently been discussed~\cite{ref7, ref8}. Indeed, skyrmions in AFMs do not suffer from demagnetizing field and are not subjected to Magnus force.

Despite being topologically stable, energetically favourable localization of skyrmions requires inclusion of additional interactions into the conventional Heisenberg Hamiltonian. Interactions leading to skyrmion stabilization are the following: long-range dipolar interactions, four-spin exchange and frustrated exchange interactions and antisymmetric Dzyaloshinskii-Moriya interaction (DMI)~\cite{ref9, ref10}. The latter one allows creation of skyrmions larger than the lattice  constant and fixes the helicity of the skyrmion. It has relativistic spin-orbit coupling nature and can arise only in the magnets with broken inversion symmetry. It was shown that skyrmions in AFMs can be stabilized solely by frustrated next-nearest neighbour interactions~\cite{ref11}. 

In the present study we consider a geometrically frustrated classical Heisenberg antiferromagnet on a triangular lattice with only nearest-neighbor interactions in the presence of the DMI and an external magnetic field. A recent study showed that at a sufficiently large strength of the DMI a three-sublattice skyrmion crystal can be stabilized within a certain range of the magnetic fields~\cite{ref12}. The goal of the present investigation is to establish the minimum values of both the DMI strength as well as the magnetic field intensity at which such a skyrmion phase can exist at very low temperatures (close to zero). 

\section{Model}
We consider the system of Heisenberg spins located at sites of a triangular antiferromagnetic lattice, the ground state of which in the absence of DMI is a planar $120^{\circ}$ configuration. In the presence of DMI and an external magnetic field the Hamiltonian can be written as

\begin{equation}
H = - J \sum_{\langle i,j \rangle}\vec{S_{i}}\cdot
\vec{S_{j}} + \sum_{\langle i,j \rangle}\vec{D_{ij}}
\cdot\Big [\vec{S_{i}}\times\vec{S_{j}} \Big] - \vec{h}\sum_i\vec{S_i},
\label{hamilt}
\end{equation}

where $\vec{S_i}$ is a classical Heisenberg vector of unit length, $J < 0$ is the AFM exchange coupling constant and $\vec{h}$ is the external magnetic field applied along $z$ direction. $\vec{D_{ij}}$ is Dzyaloshinskii-Moriya vector and its orientation is defined by the symmetries of the crystal. In our case it points along the radius-vector connecting two neighbouring sites and thus can be written in the form $\vec{D_{ij}} = D\frac{\vec{r_i} - \vec{r_j}}{|\vec{r_i} - \vec{r_j}|}$, where the parameter $D$ describes the strength of DMI.

To identify the emergence of the skyrmion phase we use a topological number for a discretized lattice as an order parameter
\begin{equation}
\chi_L = \frac{1}{4\pi N_s} \Big|\Big\langle \sum_i \Big( \chi^{12}_{i} + \chi^{34}_{i} \Big)\Big\rangle\Big|
\label{chiral}
\end{equation}

where $\chi^{ab}_{i} = \vec{S_i}\cdot[\vec{S_a}\times\vec{S_b}]$ is a chirality defined on the triangle spanned by three neighbouring spins, $\vec{S_i},\vec{S_a}$ and $\vec{S_b}$, and summation runs over one sublattice, $N_s$ being the number of sites in each sublattice, $N_s = L^2/3$. 

We use a hybrid Monte Carlo (MC) method, for which one MC sweep consists of three sequential Metropolis flips of three sublattices followed by an over-relaxation rotation of all spins. The over-relaxation method~\cite{ref13} consists of the spin rotation around the effective local field, acting on it from surroundings, in such a manner that the overall energy does not change, which allows faster decorrelation. We consider the lattice sizes of $L = 24,\ldots,72$ and use $3 - 6 \times 10^5$ MC sweeps for each temperature, including the equilibration time. For each point in the $h-D$ plane the system is gradually cooled down from higher temperatures for a proper relaxation and securing equilibrium conditions down to very low temperatures. We set $J=1$ and the Boltzmann constant $k_B=1$. 

\section{Results and conclusions}

\begin{figure}[h!]
\subfigure{\includegraphics[width=0.48\columnwidth]{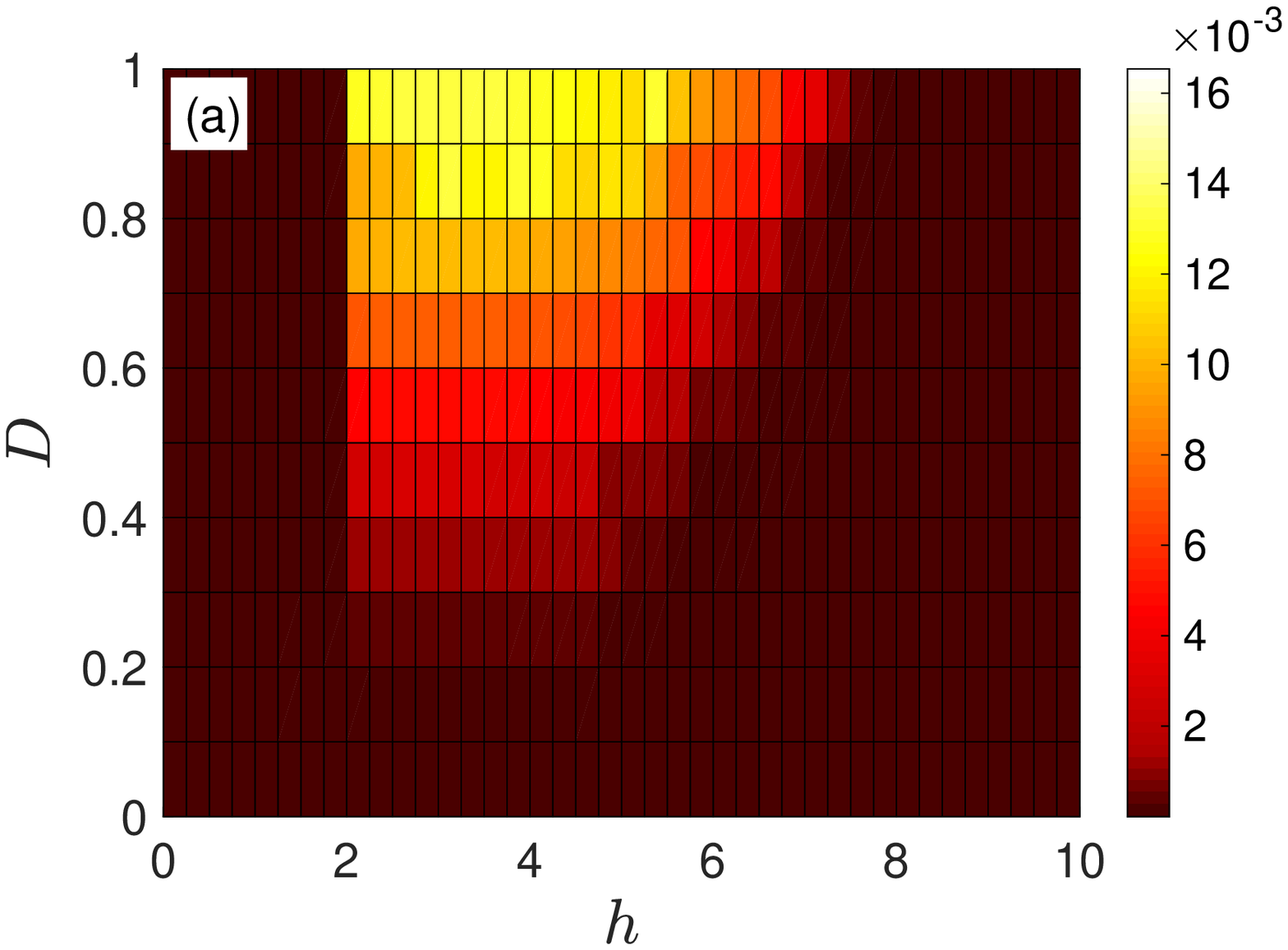}\label{fig:PD}}
\subfigure{\includegraphics[width=0.48\columnwidth]{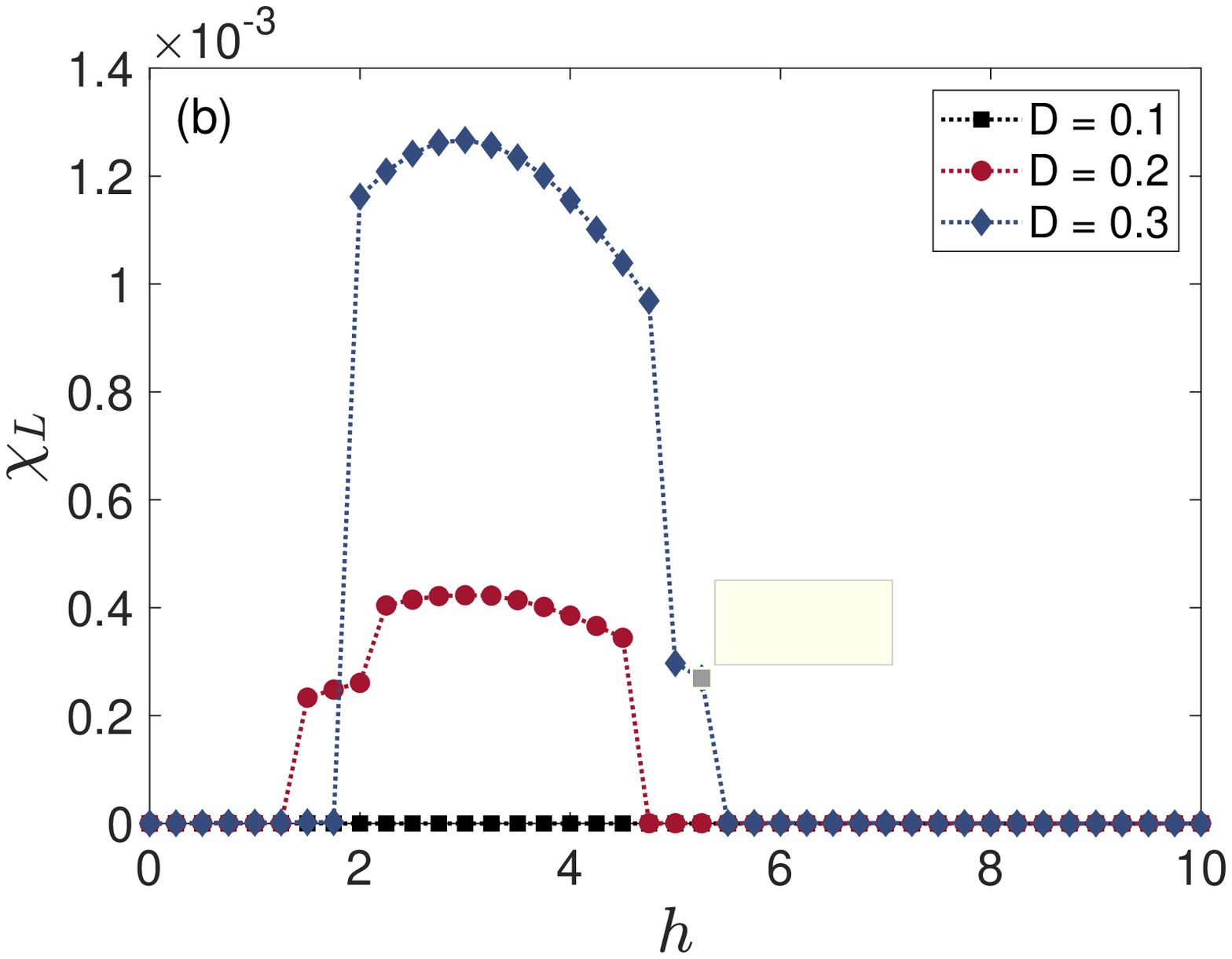}\label{fig:chir-h}}
\caption{Topological number $\chi_L$ for $T = 0.01$ and $L = 48$, shown (a) in $D-h$ plane and (b) as a function of the field for $D=0.1, 0.2$ and $0.3$.}
\label{fig:chirality}
\end{figure}

In Fig.~\ref{fig:PD} we present the topological number $\chi_L$ in $D-h$ plane, corresponding to lowest considered temperature of $T=0.01$ for $L = 48$. One can see that the region of non-zero values, which represents the skyrmion phase, shrinks with decreasing $D$ until it vanishes at $D \approx 0.2$. The parameter region of the skyrmion phase emergence/vanishing is more clearly shown in Fig.~\ref{fig:chir-h}, in which the quantity $\chi_L$ is plotted as a function of the field for three values of $D=0.1,0.2$ and $0.3$. There is clearly no skyrmion phase for $D=0.1$. For $D=0.2$ the topological number already takes relatively small but non-zero but values within $1.5\leq h \leq 2$ with a sharper increase above $h=2$ until $h=4.5$ and completely vanishing at higher fields. For $D=0.3$, there is a sharp increase of $\chi_L$ to relatively high values, which go to again zero beyond $h=5.25$ with a small step within $5\leq h \leq 5.25$.

\begin{figure}[h!]
\subfigure{\includegraphics[width=0.48\columnwidth,clip]{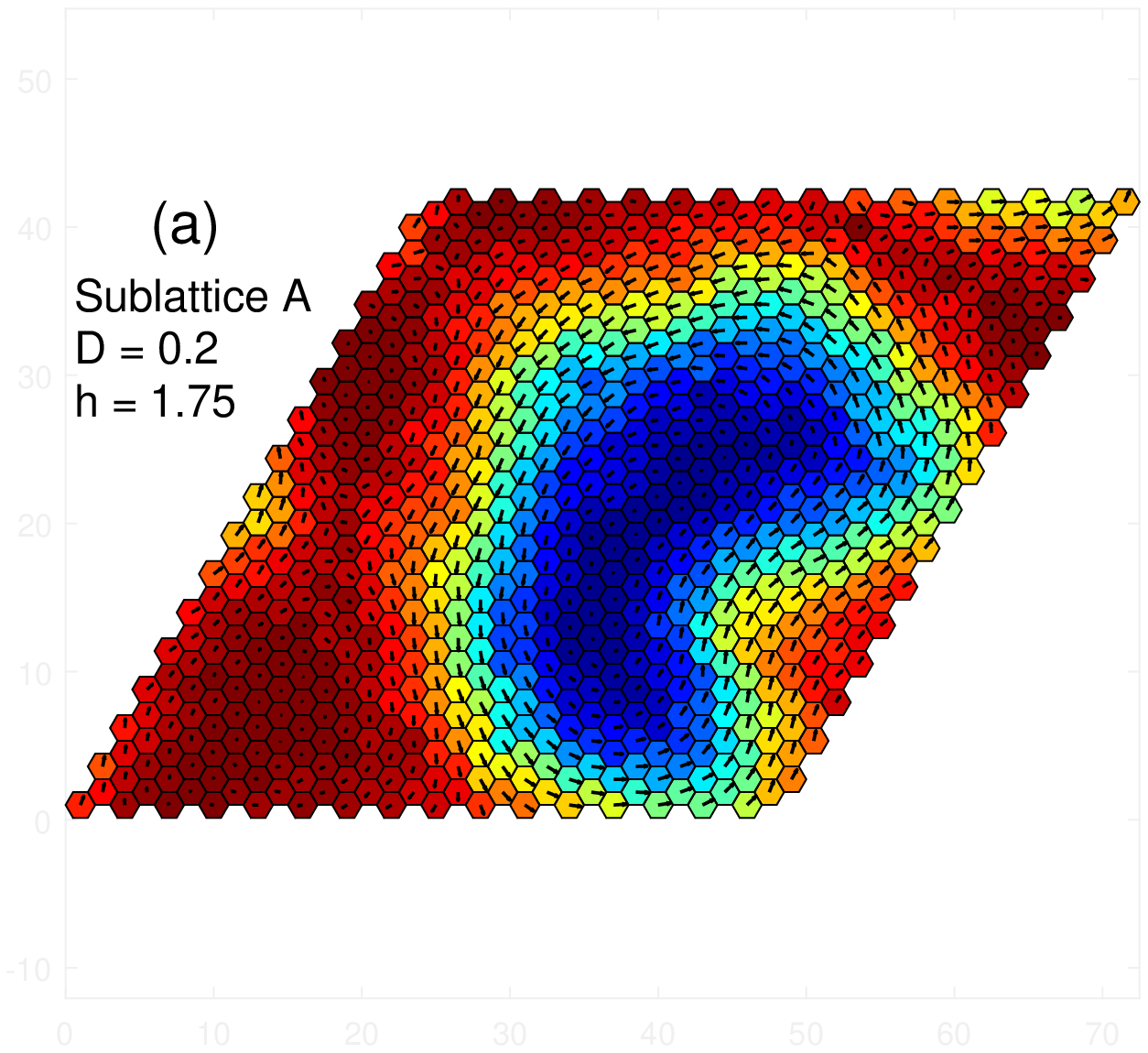}\label{fig:snap_D02_h1_75_A}}
\subfigure{\includegraphics[width=0.48\columnwidth,clip]{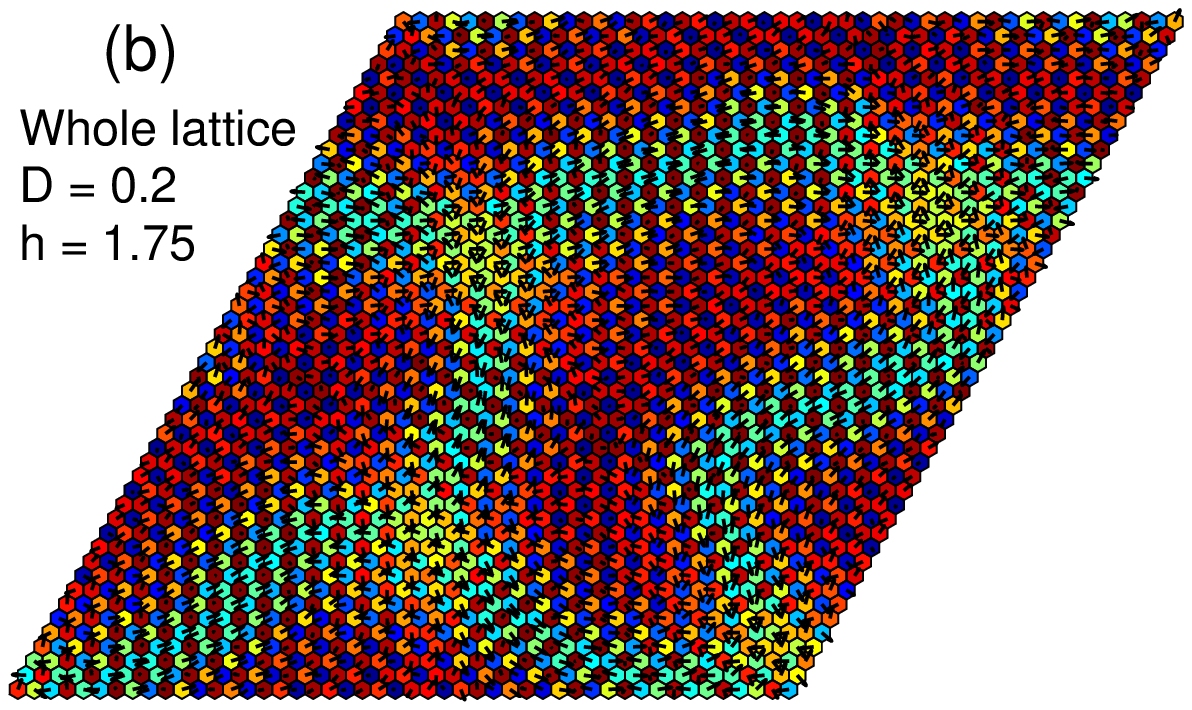}\label{fig:snap_D02_h1_75_ABC}}
\subfigure{\includegraphics[width=0.48\columnwidth,clip]{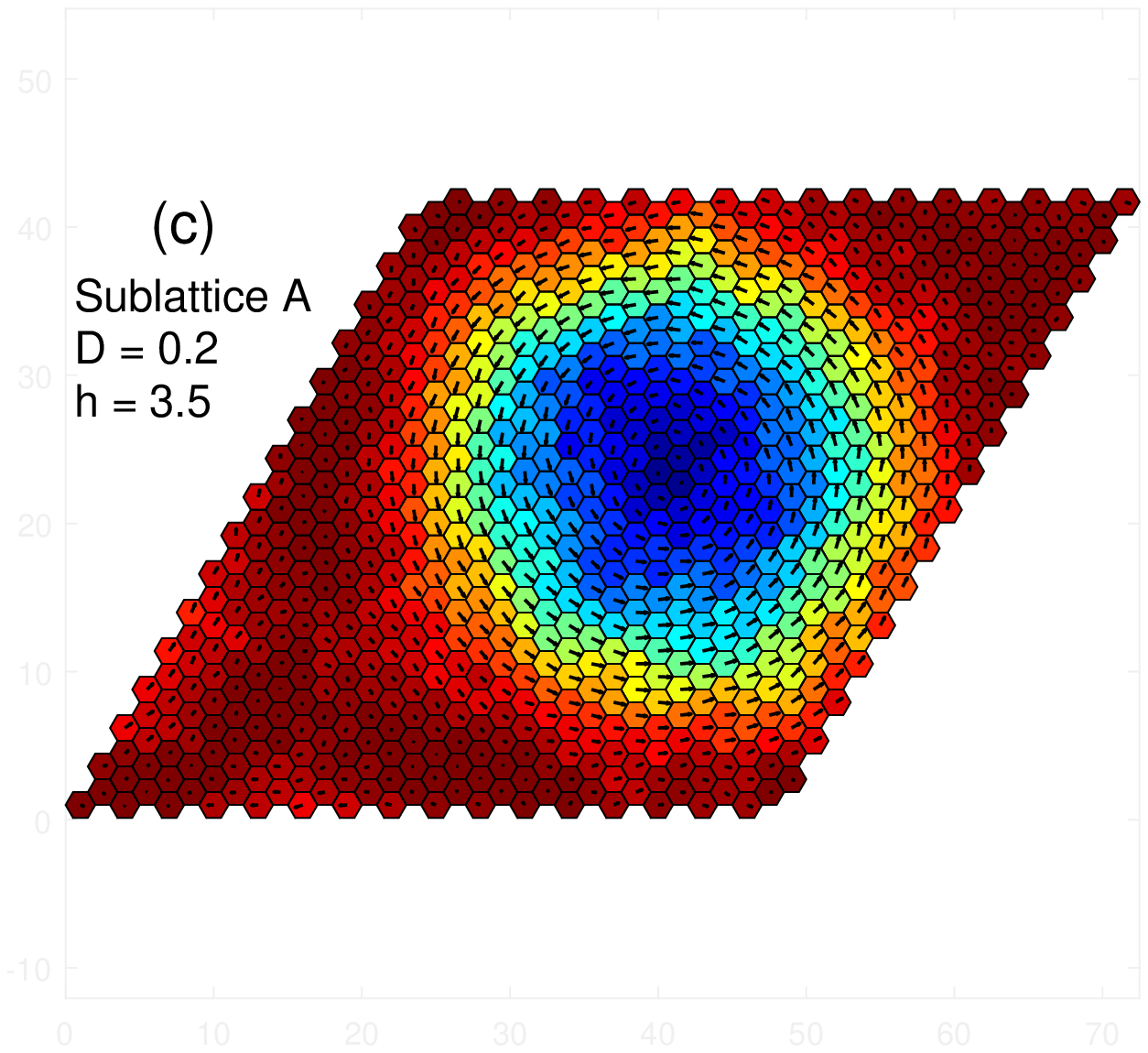}\label{fig:snap_D02_h3_5_A}}
\subfigure{\includegraphics[width=0.48\columnwidth,clip]{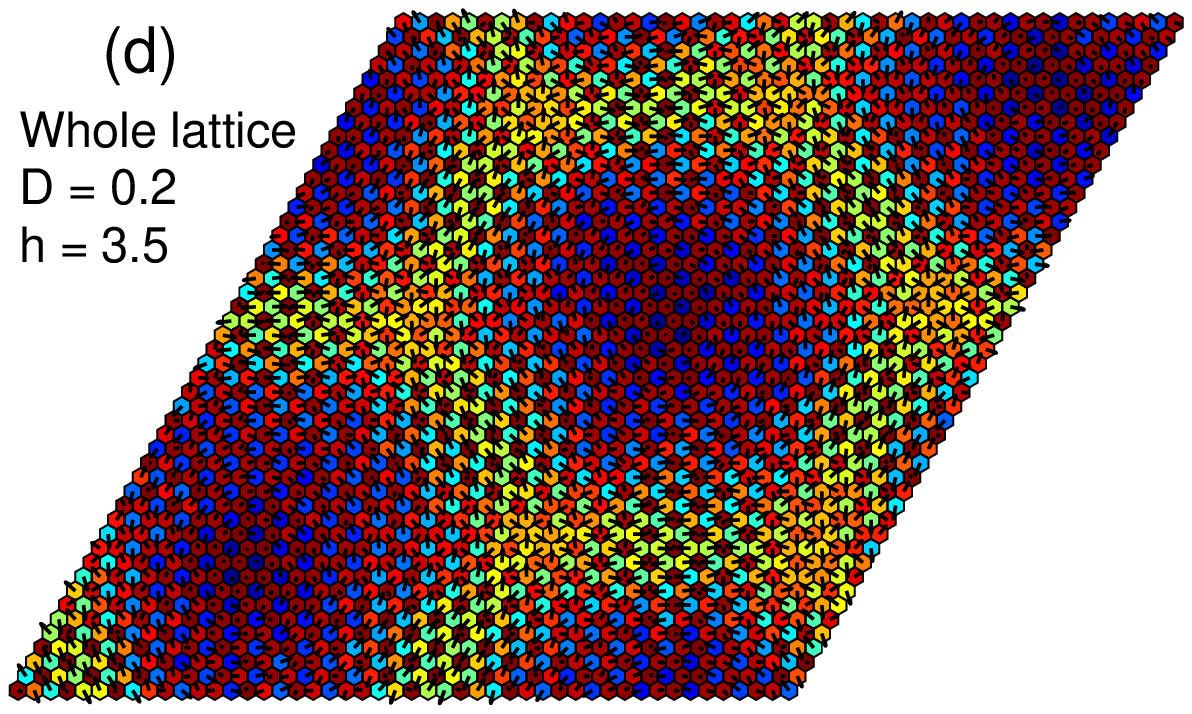}\label{fig:snap_D02_h3_5_ABC}}
\subfigure{\includegraphics[width=0.48\columnwidth,clip]{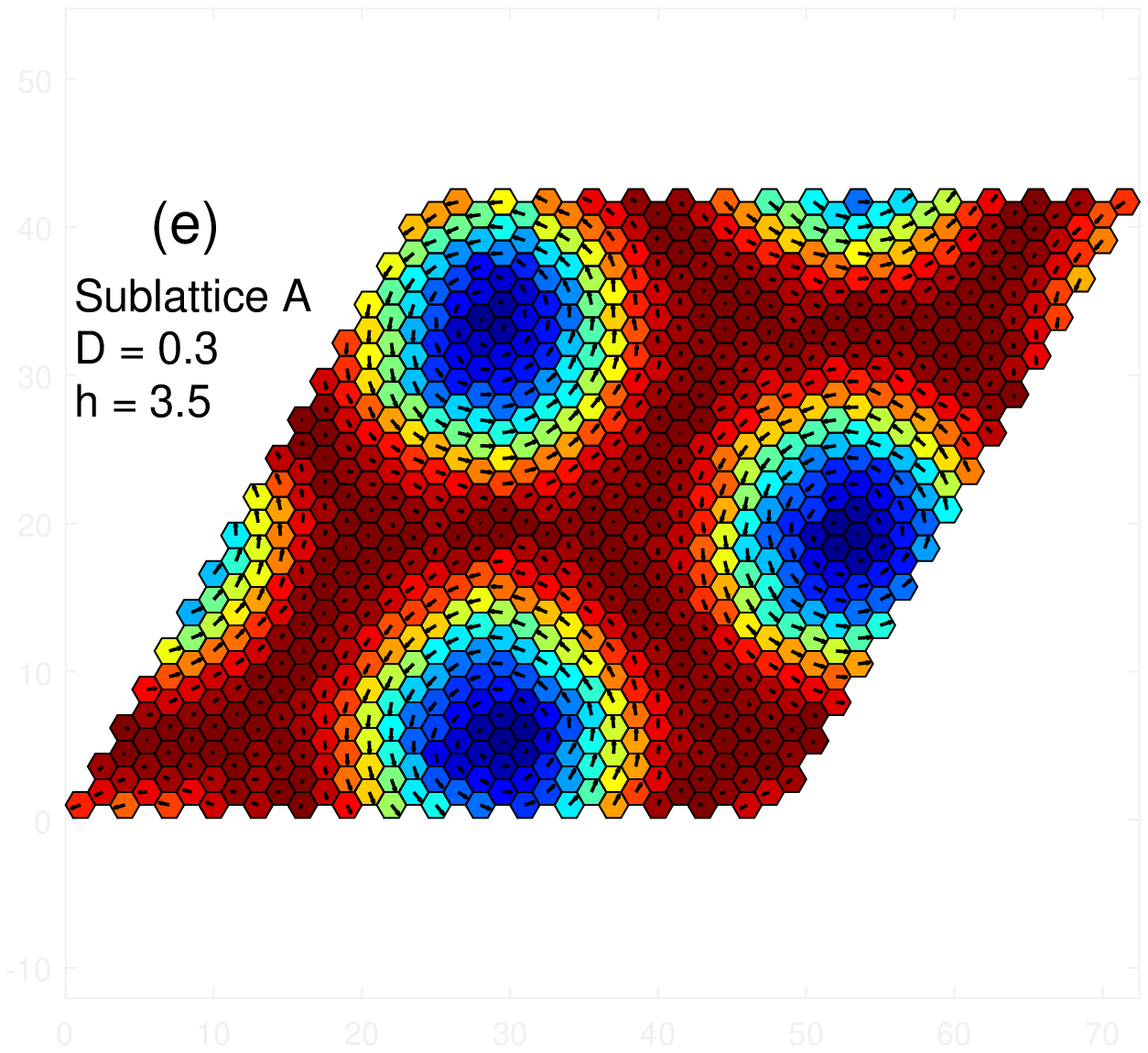}\label{fig:snap_D03_h3_5_A}}
\subfigure{\includegraphics[width=0.48\columnwidth,clip]{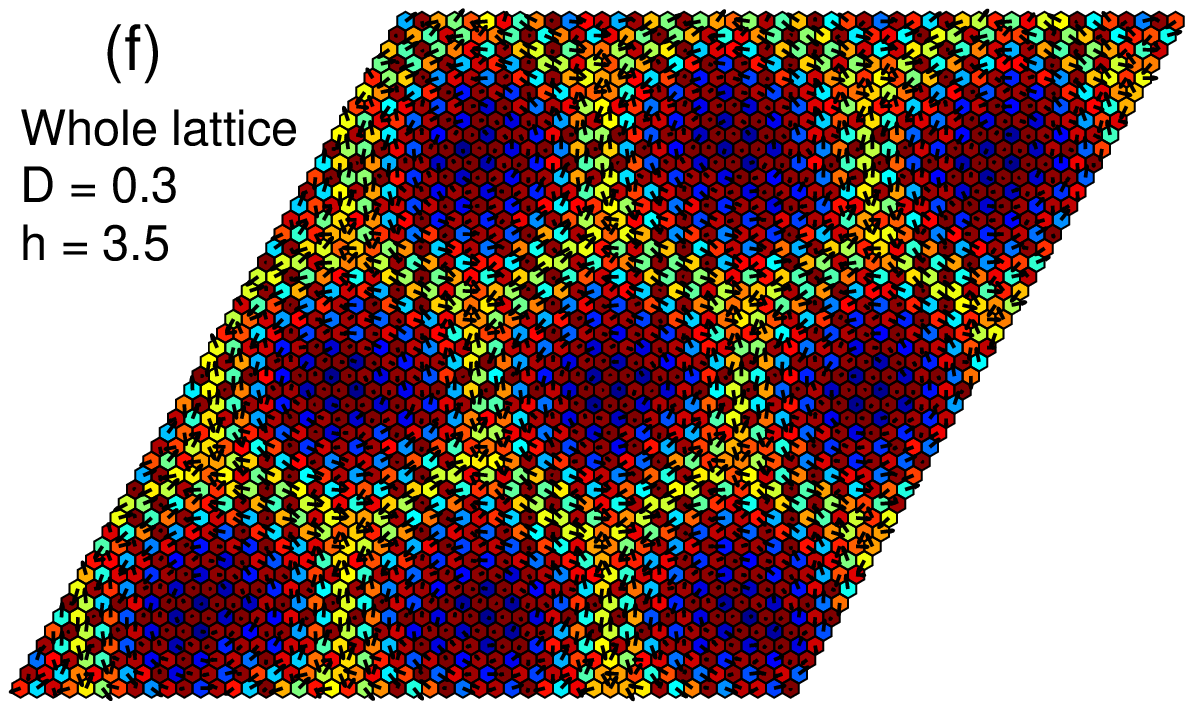}\label{fig:snap_D03_h3_5_ABC}}
\caption{Spin snapshots taken on one sublattice and the whole lattice of the size $L=48$ (only part is shown) at $T = 0.01$ and different parameter sets of $(D,h)$ equal to (a,b) $(0.2,1.75)$, (c,d) $(0.2,3.5)$, and (e,f) $(0.3,3.5)$.}
\label{fig:snaps}
\end{figure}

To verify the presence of skyrmions in the above discussed regions of the parameter space, in Fig.~\ref{fig:snaps} we show spin snapshots in one of the three sublattices (let us say A) as well as in the entire lattice. As one can observe in Figs.~\ref{fig:snap_D02_h1_75_A} and~\ref{fig:snap_D02_h1_75_ABC}, for $D=0.2$ and $1.5\leq h \leq 2$ with the small but finite values of $\chi_L$ there are some signs of the skyrmion lattice formation with a relatively large skyrmionic configuration with distorted core and irregular shape. At larger fields (see Figs.~\ref{fig:snap_D02_h3_5_A} and~\ref{fig:snap_D02_h3_5_ABC}, for $h=3.5$) the skyrmion shape becomes more circle-like but their size is still large. Further increase of the DMI parameter to $D=0.3$ at $h=3.5$ results in the formation of a skyrmion phase with smaller circular-shaped skyrmions arranged on a triangular lattice that is also typical for larger values of $D$.

In conclusion, we have established the minimal values of the parameters of DMI and the external field necessary for the emergence of the skyrmion phase in the frustrated Heisenberg antiferromagnet on a triangular lattice in near ground state conditions. The latter was identified as a region with non-zero values of the topological charge parameter. In particular, within the used resolution the skyrmion phase emerged at the DMI parameter value of $D = 0.2$ and was found to persist in the fields $2 \lesssim h < 4.5$.
\section{Acknowledgement}
This work was supported by the Scientific Grant Agency of Ministry of Education of Slovak Republic (Grant No. 1/0531/19).
%


\end{document}